\documentclass[12pt,slideColor]{iopart}
\eqnobysec
\usepackage{color}
\usepackage{amssymb}
\usepackage{graphicx}
\usepackage{dcolumn}
\usepackage{bm}
\usepackage{epsf}

\begin{document}

\jl{1}

\title{Fourier's law on a one-dimensional optical random lattice}

\author{T. Platini$^{\,1}$, R. J. Harris$^{\,2}$ and D. Karevski$^{\,3}$}

\address{$^1$ Department of Physics, Virginia Tech, Blacksburg, VA 24061, USA\\
$^2$  School of Mathematical Sciences, Queen Mary University of London, Mile End Road, London, E1 4NS, UK\\
$^3$  Institut Jean Lamour, D\'epartement Physique de la Mati\`ere et des Mat\'eriaux, Groupe de Physique Statistique, Nancy-Universit\'e CNRS,
B.P. 70239, F-54506 Vandoeuvre les Nancy Cedex, France\\
}  

\begin{abstract}
We study the transport properties of a one-dimensional hard-core bosonic lattice gas coupled to two particle reservoirs at different chemical potentials which generate a current flow through the system. In particular, the influence of random fluctuations of the underlying lattice on the stationary-state properties is investigated. We show analytically that the steady-state density presents a linear profile. The local steady-state current obeys the Fourier law $j=-\kappa(\tau)\nabla n $ where $\tau$ is a typical timescale of the lattice fluctuations and $\nabla n$ the density gradient imposed 
by the reservoirs.
\end{abstract}


\section{Introduction}
The transport properties of energy or particles in small quantum systems are an important topic in nonequilibrium  statistical dynamics. In particular, the transition between ballistic and diffusive transport is at the centre of many investigations, attempting to understand from a microscopical point of view the emergence of the celebrated Fourier law \cite{lebowitz,dhar}.
With the development of nanoscale technologies, it is now becoming possible to design proper experiments that can potentially test theoretical predictions for small quantum systems. The most promising possibilities certainly come from the optical lattice community.
For example, optical lattices can now be used to experimentally generate one-dimensional (1D) bosonic systems~\cite{moritz,tolra,kinoshita1} which have been studied theoretically for many years~\cite{girardeau,liebliniger,lenard}. In the large scattering-length limit and at low densities, ultracold bosons effectively behave as impenetrable particles~\cite{olshanii}, namely as hard-core bosons, thus realising the Tonks-Girardeau model~\cite{girardeau,liebliniger}. Experiments on such 1D hard-core bosons have been performed with Rubidium atoms within both continuum~\cite{kinoshita1} and lattice contexts~\cite{paredes}. 
For a recent review of developments in ultracold gases and optical lattices, see~\cite{bloch}.

In this letter we present the transport properties of a 1D-lattice hard-core bosonic gas driven out of equilibrium by the interaction at its boundaries with two external reservoirs that induce a particle-current flow through the system. We remark here that similar studies have been performed in \cite{michel0,michel1,dubi}.  The particular focus of the present work is the influence of lattice fluctuations, which may either be induced artificially or be inherent to an experimental set-up, on the transport properties

\section{Model}
The Hamiltonian associated with the hard-core boson model on a linear optical lattice of $N$ sites is given by
\begin{equation}
H_S=\sum_{l=1}^N h_l+\sum_{l=1}^{N-1}V_l.
\label{h1}
\end{equation}
Here the on-site one-particle Hamiltonian is
\begin{equation}
h_l=\varepsilon b^+_lb_l =\varepsilon n_l\; ,
\end{equation}
with a site-independent chemical potential $\varepsilon$ coupled to the local occupation number $n_l=b^+_lb_l$, while the hopping potential is
\begin{equation}
V_l=-t_l\left[b^+_lb_{l+1}+b_{l+1}^+b_l\right]
\end{equation}
where the hopping rate $t_l$ may depend on the position in the trap.
The creation and annihilation operators satisfy the usual bosonic commutation relations on different sites, $[b_l,b^+_{l'}]=[b_l,b_{l'}]=[b^+_l,b^+_{l'}]=0$, while the hard-core constraint is implemented by the additional conditions  $b_l^2=(b_l^+)^2=0$ and $\{b_l,b_l^+\}=1$ preventing more than single occupancy of sites.  Notice that through the transformation $b^+=(\sigma^x+\rmi \sigma^y)/2$, where $\sigma^{x,y}$ are the usual Pauli matrices, the hard-core boson Hamiltonian is exactly mapped onto the $XX$ quantum spin chain 
which, in recent years, has been studied extensively in a nonequilibrium context~\cite{karevski2}.

 The bosonic gas inside the trap is coupled at its left and right boundaries to
 ideal (non-interacting) hard-core bosonic reservoirs set at different densities $n_L$ and $n_R$, described by the single-particle density matrices
 \begin{equation}
 \rho_{L,R}=|1\rangle n_{L,R}\langle 1|+|0\rangle (1-n_{L,R})\langle 0|
 \end{equation}
where the labels $L$ and $R$ stand for the left and right reservoirs respectively and $|0\rangle,|1\rangle$ are the  associated vacuum and one-particle states. 
The interaction with the reservoirs is implemented via a discrete-time repeated interaction scheme which in the continuum limit leads to a Markovian Lindblad dynamics \cite{AttalPautrat}.
{ See \cite{wichterich} for a useful discussion on the possible failure of Lindblad dynamics in the description of stationary nonequilibrium properties and the importance of neglecting the internal couplings.} Within the discrete process, at a given time $t$ only one left reservoir particle and one right reservoir particle, in state $\rho_L$ and $\rho_R$ respectively, interact with the system.  These particles interact for a time $\tau$ through the hopping potential $V_0$ and $V_N$.  After the interaction, i.e., at time $t+\tau$, the system state $\rho_S=\Tr_E\{\rho\}$, obtained after tracing out the environment degrees of freedom corresponding to the left and right reservoirs, is given by
\begin{equation}
\rho_S(t+\tau) = \Tr_{L,R}\left\{U_{I} \left(\rho_L\otimes \rho_S(t)\otimes \rho_R\right) {U_{I}}^\dagger\right\}.
\label{rho1}
\end{equation}
Here $U_I=\rme^{-\rmi\tau H_T}$ with the total relevant Hamiltonian given by
$
H_T=H_S+V_{0}+V_N+h_{0}+h_{N+1}
$
where
$h_{0,N+1}$ are  the one-particle Hamiltonians of the reservoirs. 
Notice here that $H_T$ is of the form (\ref{h1}) with $N+2$ sites.
The process is then repeated with new reservoir particles such that (\ref{rho1}) is iterated further. 
The net effect of the process, for every timestep $\tau$, is that a boson can be injected into the trap or escape from it. 
For example, the extreme limit $n_L=0$ and $n_R=1$ describes the injection of bosons from the right of the trap and their escape to the left, i.e., in this case escape to the right and injection from the left are forbidden. 

\begin{figure}
\centerline{
\includegraphics[width=8.5cm,angle=0]{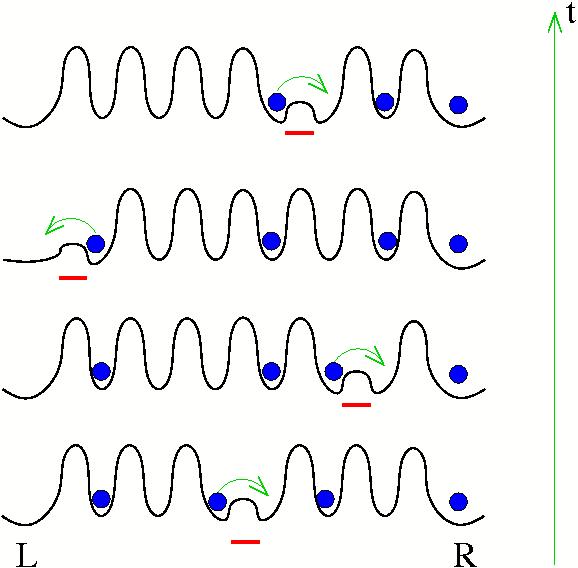}
}
\caption{\label{fig0} Sketch of the time-evolution of the fluctuating optical lattice. Here the left reservoir is empty while the right one has full occupancy. Local fluctuations, underlined by a short straight line, enhance the hopping rate locally.
}
\end{figure}

As mentioned in the introduction, we consider the effect of fluctuations of the optical lattice which may be induced by some underlying physical process, e.g., vibrations of mirrors, presence of impurities. We simply model this randomness by allowing that the  hopping rates $t_l$ $\forall l=0,...,N+1$ fluctuate in time within a typical timescale $\tau_f$. In the following we assume $\tau_f\simeq \tau$. During the time-evolution each hopping rate  follows a  stochastic trajectory, see figure \ref{fig0}, which is governed by some known probability distribution.

\section{Dynamics}
Given an initial equilibrium system state, $\rho_S(0)$, 
we start the dynamics by iterating (\ref{rho1}) with the evolution operator $U_I$ following the fluctuations of the hopping rates.  
Instead of solving directly the dynamical equation for the density matrix, we 
study the time-evolution of correlation functions. 
Moreover, due to the free-fermionic structure of the model after a Jordan-Wigner transformation  \cite{lieb} , 
\begin{equation}
\begin{array}{l}
{\Gamma}_{l}={A}_{l}=\rme^{\rmi\pi\sum_{j=1}^{l-1}n_j}(b_l+b_l^+)\\
{\Gamma}_{N+2+l}=-i{B}_{l}=-\rmi\rme^{\rmi\pi\sum_{j=1}^{l-1}n_j}(b_l-b_l^+)
\end{array}
\forall l=0,...,N+1
\end{equation}
where the $\Gamma$s are Majorana real (Clifford) operators satisfying ${\Gamma}^\dagger=\Gamma$ and $\{{\Gamma}_i,{\Gamma}_j\}=2\delta_{ij}$,
{ thanks to Wick's theorem, one can express all physical observables in terms of the two-point correlation functions}
\begin{equation}
\left[G(t)\right]_{jk}=\frac{\rmi}{2}\Tr\left\{ \big[{\Gamma}_k,{ \Gamma}_j\big]\rho(t)\right\}
	\; .
	\label{g1}
\end{equation}  
In the Heisenberg picture, the time evolution of the Majorana field ${\Gamma}$, generated by $H_T$, is simply given by ${ \Gamma}(t)=\rme^{-\rmi tT}{\Gamma}(0)\equiv { R}(t){ \Gamma}(0)$, where $T$ is defined by the Hamiltonian in terms of the field $\Gamma$:
$
H_T=(1/4) {{ \Gamma}}^\dagger T {\Gamma}
$. The matrix elements of the rotation matrix $R$ are simply expressed in terms of the spectral properties of $H_T$, see~\cite{karevski1} for the explicit forms.

We order the ${{\Gamma}}^\dagger=({ \Gamma}^\dagger_E,{ \Gamma}^\dagger_S)$ such that the first part, ${ \Gamma}_E$, is associated with the interacting part of the environment and the second part, ${ \Gamma}_S$, with the components of the system. Projecting (\ref{g1}) onto the system part, one arrives at the fundamental dynamical equation for the system correlation matrix $G_S$:
\begin{equation}
G_S(t+\tau)={ R}_S G_S(t){{ R}_S}^\dagger+{ R}_{SE} G_E{{ R}_{SE}}^\dagger \; .
\label{eqfond}
\end{equation}
The $2N\times 2N$ matrix  $R_S$ is that part of the full rotation matrix 
$
R=\rme^{-\rmi\tau T}= \left(\begin{array}{cc}
R_E&R_{ES}\\
R_{SE}&R_S
\end{array}\right)
$ 
which acts on the system. The $2N\times 4$ rectangular matrix $R_{SE}$ is given by the lower off-diagonal block of $R$ expressed in the basis $({ \Gamma}^\dagger_E,{ \Gamma}^\dagger_S)$.  For non-interacting dynamics, i.e., a closed system,  $R_{SE}=0$ and the rotation matrix splits into a block-diagonal form where $R_S$ and $R_E$ are the rotation matrices of the system and environment part respectively.
In the dynamical equation (\ref{eqfond}), the bath properties enter only through the initial environment particle states, encoded in the two-point correlation matrix $G_E$. The correlation matrix $G_E$ stays constant in time since at each step of the repeated interaction procedure the bath particles are replaced by  fresh ones.

\section{Steady-state current and density}
In the following, we concentrate mainly on the asymptotic properties of (\ref{eqfond}). The steady state is reached exponentially with a relaxation time depending on the system size N. For the non-disordered situation the timescale needed to reach the steady state behaves as $N^3$ \cite{prosen}.  
In particular, we focus our attention on the transport properties of the bosonic gas through the optical trap.  For that we compute the density profile and  the particle current along the chain.  Since the hopping dynamics conserves particles, one may naturally define the particle current through the Heisenberg equation of motion for the density: 
\begin{equation}
\dot{\hat{n}}_l=\rmi[H_S,\hat{n}_l]\equiv {\cal J}_{l-1}-{\cal J}_l
\end{equation} 
where ${\cal J}_l$ denotes the particle-current operator associated with the $l$th bond and given by 
\begin{equation}
{\cal J}_l\equiv t_l { J}_l=\rmi t_l\left[b_lb_{l+1}^+-b_l^+b_{l+1}\right]\; .
\end{equation}
The density $\hat{n}_l$ and the current $J_l$, are easily expressed in terms of the Majorana field $\Gamma$ and their expectation values $n_l=\langle \hat{n}_l\rangle =\Tr\{\hat{n}_l \rho\}$ and $j_l=\langle J_l\rangle=\Tr\{J_l \rho\}$
are given by on-site and two-site two-point correlation functions $G_S$: $n_l=(1-(G_S)_{l,l+N})/2$ and $j_l=(G_S)_{l,l+1}$. In the following the stared quantities $n^*$ and $j^*$ have to be understood as expectation in the steady-state.

\subsection{Non-fluctuating lattice}
If the lattice is free of any disorder, the hopping rates are uniform, i.e., $t_l=t_o$ $\forall l$. In this case, an exact solution of the steady state has been given in \cite{platkar}. It is found that the density profile is flat except at sites $1$ and $N$ which are directly in contact with the reservoirs. The density in the flat region is given by the mean value set by the reservoirs  $n^*_l=\bar{n}=(n_L+n_R)/2$ $\forall l\ne 1,N$ while the boundary values are 
$n^*_1=\bar{n}  - \Delta(t_o)  (n_{R}-n_{L}) / 2$
and
$n^*_N=\bar{n} + \Delta(t_o)  (n_{R}-n_{L}) / 2$ with a shift from the mean value $\bar{n}$ depending on the density difference $n_{R}-n_{L}$ and where $\Delta(t_o)=\frac{ \gamma^2}{  1 +  \gamma^2 }$ with $\gamma=t_o/2$. The steady-state current $j^*$ takes a constant value 
$j^*= -\alpha( n_{R}-n_{L} )$
independent of the system-size where $\alpha$ is a non-monotonous function of the hopping rate $t_o$, with a maximum current state at $t_o=2$. The size-independence of $j^*$ signals the ballistic nature of the transport, which is ultimately related to the integrability of the model, that is to the equations of motion of the free quasiparticles describing the system.   In this case there is no finite conductivity $\kappa$ and the system obviously does not obey Fourier's law. One may notice that this behaviour is very similar to the behaviour observed in the classical Reider-Lebowitz-Lieb model of a homogeneous harmonic chain \cite{reider} in contact with stochastic heat baths at the boundaries.

\subsection{Fluctuating lattice}
Next we study the effect of fluctuations of the lattice parameters on the steady-state properties.  We consider local fluctuations in the sense that within the timescale $\tau$ of the fluctuation only one bond is affected leading to an enhancement of the local hopping rate from its unperturbed value $t_o$ to a larger value $t_D$, which we choose to be $1/2$. Moreover, we take the limit of a strongly localized lattice gas, with $t_o\ll 1$. In other words, at each timestep $\tau$ of the dynamics, a single bond is activated at random and locally the particles are exchanged with a rate $t_D=1/2$, either within the system if the selected bond is a system one or with the reservoirs if the fluctuations act close to the boundaries.  These particle exchanges are reminiscent of the well-studied symmetric exclusion process~\cite{liggett}; the dynamics~(\ref{eqfond}) leads to a non-trivial dependence of the bulk density gradient on the interaction time $\tau$, as we shall now see.

In the weak-hopping limit $t_o\rightarrow 0$, the dynamics simplify considerably since at a given time step only one bond is activated. 
The time evolution of the system is most simply expressed in terms of Dirac fermions $c_k=(A_k+B_k)/2$ and $c_k^+=(A_k-B_k)/2$. One has in matrix form $c(t+\tau)=e^{itK}c(t)$ where $K$ is the coupling matrix defining the Hamiltonian $H_T=-c^+Kc$ in terms of the fermi field $c^+=(c_0^+,c_1^+,\dots,c_N^+,c_{N+1}^+)$. Due to the fact that at a given time step only one bond is non-vanishing, suppose bond $l$ conecting the sites $l$ and $l+1$, one has the trivial dynamics $c_k(t+\tau)=e^{it\varepsilon}c_k(t)$ $\forall k\neq(l,l+1)$ and 
$
c_l(t+\tau)=e^{it\varepsilon}[\cos\frac{\tau}{2} c_l(t)+i\sin \frac{\tau}{2} c_{l+1}(t)]
$ and
$
c_{l+1}(t+\tau)=e^{it\varepsilon}[i\sin\frac{\tau}{2} c_l(t)+\cos \frac{\tau}{2} c_{l+1}(t)]
$ 
for the on bond operators.

Consider the density gradient $\delta_l\equiv \langle n_{l+1}\rangle -\langle n_{l}\rangle$ on bond $l$. This gradient is changed only if the activated bond is the $l$th one or one of the nearest neighbours $l-1$ and $l+1$.
From the dynamical equation (\ref{eqfond}) in terms of the fermi operators, if the hopping is enhanced on bond $l$ then 
$\delta_l$ is mapped to 
\begin{equation}
\delta_l'=\delta_l \cos \tau + j_l \sin \tau
\end{equation}
and $j_l$ is mapped to
\begin{equation}
j_l'=-\delta_l \sin \tau + j_l \cos\tau\; .
\end{equation}
Alternatively, if the activated bond is $l\pm 1$, the updated density gradient satisfies
\begin{equation}
\delta_l'=\delta_l-\frac{1}{2}(\delta_{l\pm1}\cos\tau+j_{l\pm1}\sin\tau-\delta_{l\pm1})
\end{equation}
while the updated current is given by
\begin{equation}
j_l'=j_l\cos \frac{\tau}{2}+P_{l}
\end{equation}
where the $P_{l}$ are proportional to correlations across two bonds. 
As mentioned before, under this dynamics the system relaxes exponentially towards a current carrying steady state.  Nevertheless, we remark here that the periodicity of the dynamics implies a strong slowing-down of the relaxation to the steady state in the neighbourhood of $\tau=n2\pi$.  Indeed, for even values of $n$ the dynamical generator maps to the identity and for odd values it maps to a reflection dynamics which loses its relaxation properties.  In the following analysis we avoid these special $\tau$ values.

\begin{figure}
\centerline{
\includegraphics[width=9cm,angle=0]{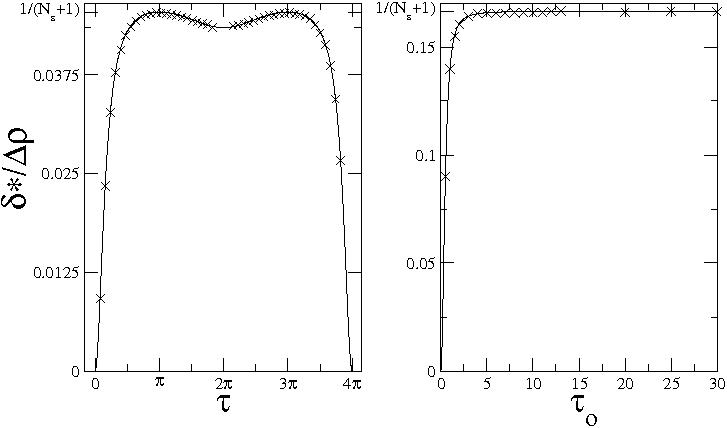}
}
\caption{\label{fig1} Normalized steady-state density gradient as a function of the interaction time with a delta time-distribution on the left and an exponential one with mean $\tau_o$ on the right. The full lines correspond to the analytical curves while the crosses are obtained numerically with a time average of the density gradient.
}
\end{figure}

In the steady state, the $P$ terms vanish on average and the set of dynamical equations for the gradient density and current closes. At any given time, the last update on bond $l$ has probability $1/3$ to have resulted from activation on bond $l$, probability $1/3$ to have resulted from activation on bond $l-1$, and probability $1/3$ to have resulted from activation on bond $l+1$. Consequently, the steady-state average (denoted by a star) gradient obeys
\begin{equation}
\delta^*_l(\cos\tau -1)+j^*_l\sin\tau=\eta(\tau) \label{ssgrad}
\end{equation}
where $\eta(\tau)$ is a constant independent of the bond index $l$. Since the steady-state current $j^*_l=j^*$ is constant in space it also follows from~(\ref{ssgrad}) that the gradient density is site-independent and we can thus omit the $l$-subscripts.
The steady-state current satisfies
\begin{equation}
j^*=\frac{1}{3}\left[-\delta^*\sin\tau+ j^*\cos\tau \right]+\frac{2}{3}j^*\cos\frac{\tau}{2}
\end{equation}
which is equivalent to 
\begin{equation}
j^*=-\frac{\sin\tau}{3-\cos\tau-2\cos\frac{\tau}{2}}\delta^*\equiv - \kappa(\tau)\delta^*
\end{equation}
and defines the conductivity coefficient $\kappa(\tau)$.
We now determine the constant $\eta(\tau)$ by considering the boundary conditions. Remembering that the densities on the reservoir sites are fixed and that the boundary terms $j_0$ and $j_N$, which are initial correlations between the reservoirs and the system,  vanish due to the repeated interaction scheme, one sees that an update on bond $0$ (between left reservoir and boundary site) gives $\delta'_0=\frac{1}{2}\delta_0(1+\cos\tau)$ whereas an update on bond $1$ gives
$\delta_0'=\delta_0-\frac{1}{2}(\delta_1\cos\tau+j_1\sin\tau-\delta_1)$.
The steady-state average  $\delta_0^*$ must therefore obey
\begin{equation}
\delta_0^*=\frac{1}{2}\left(\delta_0^*\frac{1+\cos\tau}{2}\right)+\frac{1}{2}\left(
\delta_0^*-\frac{\delta^*\cos\tau+j^*\sin\tau-\delta^*}{2}
\right)
\end{equation}
giving 
$
\eta(\tau)=(\cos \tau-1)\delta_0^*\;.
$
By symmetry, at the right-boundary we find $\delta^*_N=\delta^*_0$. Noting that the reservoir density difference $\Delta n\equiv n_R-n_L=2\delta_0^*+(N-1)\delta^*$, one finally gets for the bulk steady-state density gradient
\begin{equation}
\delta^*=\frac{\Delta n}{N+1+\gamma(\tau)}
\end{equation}
with the finite-size shift function (extrapolation length) 
\begin{equation}
\gamma(\tau)=2\frac{\sin\tau}{1-\cos\tau}\kappa(\tau)
\end{equation}
This analytical expression is compared with numerical simulation data in figure \ref{fig1} obtained on chains of $N=30$ spins and the agreement is seen to be excellent.

So far, we have considered the somewhat unphysical situation where the enhancement of a local hopping rate always stays precisely for a time $\tau$. This hypothesis leads to the trigonometric form of the conductivity $\kappa(\tau)$ and shift function $\gamma(\tau)$.  A more reasonable assumption would be to draw the duration of a local fluctuation from a probability distribution $f(\tau)$. In that case, one may average the dynamical equations for the current and the gradient density over the time-distribution $f(\tau)$ which basically leads to replacing the trigonometric functions $\cos\tau$, $\sin\tau$ and $\cos \frac{\tau}{2}$ by their expectations under $f$. For example, for an exponential distribution of interaction timescales with mean $\tau_o$, one gets the shift function $\gamma=(2/\tau_o) \kappa$ with conductivity $\kappa=\frac{\tau_o^2+4}{3\tau_o(\tau_o^2+2)}$ which is again in agreement with the numerical results, see figure~2.
In the limit of short-time interaction $\tau_o\rightarrow 0$, conductivity diverges as $\tau_o^{-1}$ while the density gradient goes as $\tau_o^2$.  This leads to a linear vanishing of the steady-state current $j^*\sim \tau_o$.

If lattice fluctuations are generated diffusively, e.g., the updated bond follows a symmetric random walk, then analysing the dynamical equations along the same lines, we obtain again a linear profile of the particle density in the steady state: $\delta^*=\Delta\rho/(N+1+\gamma(\tau))$ where the shift function $\gamma(\tau)$ depends on the precise definition of the random walk at the boundaries.

At finite, but small, unperturbed hopping rates $t_l=t_o$ along the optical lattice, we observe numerically that the linear density profile survives. However, the density gradient is strongly attenuated by a function of the bulk hopping rate $t_o$, whose asymptotic behaviour is $(N t_o)^{-1}$ for large lattice sizes.  Consequently, for large systems the steady-state density gradient behaves as $\delta^*\sim1/N^2$ instead of the former $1/N$ behaviour. At the same time, the steady-state current is $j^* \sim 1/N$ which leads to a linear divergence of the conductivity coefficient with system size: $\kappa\sim j^*/\delta^*\sim N$. This implies that the classical transport properties of the hard-core boson gas are induced by the optical lattice fluctuations only for sufficiently small $Nt_o$ values. In the thermodynamical limit, $N\rightarrow \infty$, the system shows a ballistic transport behaviour. See~\cite{michel1} for a similar discussion.

\section{Conclusion} 
We have derived analytical expressions for the steady-state conductivity and density profile of a nonequilibrium hard-core boson model. For a perfect optical lattice, due to the integrability of the model, the transport properties are anomalous with an infinite conductivity coefficient, reflecting the ballistic nature of the excitations. On the other hand, when fluctuations of the underlying lattice are present and when $Nt_o$ is sufficiently small, the classical Fourier law is recovered, with  a linear density profile and a finite conductivity coefficient depending on the lattice fluctuation properties.

Fourier's law in nature is so wide-spread that specific choices of model parameters should not play a decisive role in deriving it. Consequently, we do not believe that the dynamical fluctuations considered in the present paper have to be the generic origin of normal heat conduction. In general it is believed that the basic microscopic mechanism leading to Fourier's law is linked to the scattering of energy carriers, inducing mixing properties. Indeed, since the thermal conductivity, obtained from the Green-Kubo formula, is given by an infinite integral of the autocorrelation function of the current operator, this correlator has to decay quickly enough such that the integral converges even in the thermodynamic limit.
In the present work we have presented a possible way of how Fourier's law can be established in some limiting situations by introducing dynamical fluctuations which somehow scatters the energy carriers. The origin of such fluctuations at the microscopic level can be of course very diverse,  as for example a coupling to phonon modes or external perturbations like vibrations of the mirors in an optical setup. 
The next step of this work will be the study of the current autocorrelation function to compare our results for the conductivity coefficient with the Green-Kubo expectation, see for example \cite{saito} for a study in this spirit.

\emph{Acknowledgements:} RJH thanks the L'institut Jean Lamour and the Universit\'e Henri Poincar\'e, Nancy for kind hospitality.  We are grateful to the Nancy Groupe de Physique Statistique, to St\'ephane Attal and to the referees for useful discussions and comments. 
Ce travail a b\'en\'efici\'e d'une aide de l'Agence Nationale de la Recherche  
portant la r\'ef\'erence ANR-09-BLAN-0098-01. T. P. has been supported by the U.S. National Science Foundation through DMR-0705152.

\end{document}